\documentclass{jfm}
\usepackage[T1]{fontenc}
\usepackage[latin9]{inputenc}
\usepackage{verbatim}
\usepackage{soul}
\usepackage{xcolor}
\usepackage{graphicx}
\usepackage{stmaryrd}
\usepackage{amssymb}
\usepackage{amsmath}
\usepackage{esint}
\usepackage{units}
\usepackage{wasysym}
\usepackage[authoryear]{natbib}
\PassOptionsToPackage{normalem}{ulem}
\usepackage{ulem}

\makeatletter

\providecolor{lyxadded}{rgb}{0,0,1}
\providecolor{lyxdeleted}{rgb}{1,0,0}

\def\vec#1{\boldsymbol{#1}}
\def\tilde#1{\boldsymbol{\mathrm{#1}}}
\def\underline#1{\boldsymbol{\mathsf{#1}}}

\makeatother

\begin{document}

\title[Fractality of metal pad instability threshold]
{Fractality of metal pad instability threshold\\ in rectangular cells}

\author[G. Politis, J. Priede]{Gerasimos Politis \and~J\={a}nis Priede}
\affiliation{Fluid and Complex Systems Research Centre,\\
Coventry University, UK}

\maketitle

\begin{abstract}
We analyse linear stability of interfacial waves in an idealised model
of an aluminium reduction cell consisting of two stably stratified
liquid layers which carry a vertical electric current in a collinear
external magnetic field. If the product of electric current and magnetic
field exceeds a certain critical threshold depending on the cell design,
the electromagnetic coupling of gravity wave modes can give rise to
a self-amplifying rotating interfacial wave which is known as the
metal pad instability. Using the eigenvalue perturbation method, we
show that, in the inviscid limit, rectangular cells of horizontal
aspect ratios $\alpha=\sqrt{m/n}$, where $m$ and $n$ are any two
odd numbers, can be destabilised by an infinitesimally weak electromagnetic
interaction while cells of other aspect ratios have finite instability
thresholds. This fractal distribution of critical aspect ratios, which
form an absolutely discontinuous dense set of points interspersed
with aspect ratios with non-zero stability thresholds, is confirmed
by accurate numerical solution of the linear stability problem. Although
the fractality vanishes when viscous friction is taken into account,
the instability threshold is smoothed out gradually and its principal
structure, which is dominated by the major critical aspect ratios
corresponding to moderate values of $m$ and $n$, is well-preserved
up to relatively large dimensionless viscous friction coefficients
$\gamma\sim0.1$. With a small viscous friction, the most stable are
cells with $\alpha^{2}\approx2.13$ which have the highest stability
threshold corresponding to the electromagnetic interaction parameter
$\beta\approx4.7$.
\end{abstract}

\section{Introduction}

Stably stratified liquid layers carrying strong vertical electric
current are the core element of Hall--H\'eroult aluminium reduction
cells \citep{Evans2007} and the recently invented liquid metal batteries
\citep{Kim2013}. It is well known that an aluminium reduction cell
may become unstable when the depth of electrolyte layer is decreased
below a certain critical threshold which depends on the current strength
and the cell design. The metal pad instability, as it is generally
known \citep{Urata1985}, is caused by a rotating interfacial wave
\citep{Sele1977} and manifests itself as a sloshing of liquids in
the cell \citep{Davidson2000}. If the sloshing becomes too strong,
it can disrupt the operation of the cell by creating a short circuit
between the molten aluminium and the carbon anode, which is immersed
in the electrolyte at the top. There are concerns that metal pad instability
could affect also liquid metal batteries once they reach a sufficiently
large size \citep{Kelley2018,Horstmann2018,Tucs2018,Tucs2018a,Molokov2018}.
In contrast to the Tayler (internal pinch) instability, which is driven
by the interaction of electric current with its own magnetic field
and could also affect large-size liquid metal batteries \citep{Stefani2011,Weber2015,Herreman2015,Priede2017},
the metal pad instability is caused by the presence of an external
magnetic field, in particular, its vertical component, in the cell.

The mechanism behind rotating interfacial waves is clear. Firstly,
as the electric current flows through the poorly conducting electrolyte
layer following the path of least resistance, a slight initial tilt
of the interface redistributes current density from the depressed
to the elevated parts of the interface. Secondly, in order to leave
the system uniformly through the relatively poorly conducting carbon
cathode, the current spreads out horizontally through the aluminium
layer. As the horizontal electric current flows across the vertical
magnetic field, a transverse electromagnetic force arises which tries
to tilt the interface cross-wise to its initial slope so making it
rotate.

This basic mechanism is captured by the experimental model of \citet{Pedchenko2009}.
Although the same mechanism is reproduced by the electromechanical
pendulum model of \citet{Davidson1998}, which exhibits a roll-type
instability when the frequencies of two perpendicular oscillation
modes occur sufficiency close, this model misses the fact that the
metal pad instability is essentially a boundary effect \citep{Lukyanov2001}.
Namely, in the basic instability model, the electromagnetic force
is mathematically reducible to a boundary condition which describes
electromagnetically modified reflection of interfacial gravity waves
by the sidewalls. 

The simplest theoretical model of the metal pad instability is due
to \citet{Lukyanov2001} who considered a semi-infinite domain (half-plane)
bounded by a single lateral wall. They studied the effect of electromagnetic
force on the reflection of interfacial gravity waves from the sidewalls,
but missed the metal pad instability itself. This instability was
first identified in the semi-infinite model by \citet{Morris2003}
and then revisited by \citet{Molokov2011}. In this model, the interface
can be destabilised by an arbitrary weak electromagnetic effect which
gives rise to a nearly transverse wave travelling along the wall.
In the infinite channel bounded by two parallel walls, which was considered
first by \citet{Morris2003}, a sufficiently strong electromagnetic
effect is required to destabilise the interface. In this model, the
instability emerges as a longitudinal standing wave. These two basic
models are reviewed and analysed in more detail in appendix \ref{sec:appA}.

The threshold of the metal pad instability in realistic finite-size
cells is expected to depend on their geometry. For example, the square
\citep{Bojarevics1994} and cylindrical \citep{Davidson1998,Lukyanov2001}
cells are known to be inherently unstable; like the semi-infinite
cell, they can be destabilised by a relatively weak electromagnetic
effect. The low stability threshold of the square cell is attributed
to the occurrence of several pairs of gravity wave modes with the
same frequency \citep{URATA1976}. The same applies also to cylindrical
cells, which are considered as an option for the liquid metal batteries
\citep{Herreman2019,Horstmann2019}. According to the electromechanical
model of \citet{Davidson1998}, such cells are expected to be inherently
unstable. However, their numerical results as well as those of \citet{Sneyd1994}
indicate that this is not in general the case as some rectangular
cells with degenerate gravity wave spectra appear to have significantly
higher stability thresholds than the square cell. Unfortunately, \citet{Sneyd1994}
did not notice this key result in their low-resolution numerical solution,
which displays a very irregular variation of the instability threshold
with the cell aspect ratio. Like \citet{Bojarevics1994} they claim
all aspect ratios where two natural gravity wave frequencies coincide
to be critical. Although the same result is predicted by the electromechanical
pendulum model of \citet{Davidson1998}, their numerical results obtained
using the shallow-water model reveal that the gravity wave modes with
the closest frequencies are not always the most unstable ones in rectangular
cells. They explain this by the observation that only one in approximately
five mode pairs are actually coupled and thus could become unstable.

Despite the relatively long history of this problem, it is not yet
clear how the aspect ratio of rectangular cells affects their stability.
This is a question of fundamental as well as practical importance
which is addressed in the present study by carrying out a comprehensive
theoretical and numerical analysis of the basic aluminium reduction
cell model.

We show that the degeneracy is necessary but not in general sufficient
for the instability because not all degenerate gravity-wave modes
are electromagnetically coupled. The critical aspect ratios predicted
by the basic metal pad instability model are found to be limited to
$\alpha=\sqrt{m/n},$ where $m$ and $n$ are any two odd numbers.
These unstable aspect ratios form an absolutely discontinuous dense
set of points which intersperse aspect ratios with finite (non-zero)
stability thresholds. Although the fractality vanishes when viscous
friction is taken into account, the instability threshold is smoothed
out gradually and preserves its principal structure up to relatively
high values of viscous friction.

The paper is organised as follows. In the next section, we formulate
the problem and introduce a fully nonlinear two-layer shallow-water
approximation which involves a novel integro-differential equation
for the electric potential. Section $\S$\ref{subsec:cell} presents
linear stability analysis of rectangular cells which is carried out
analytically using an eigenvalue perturbation method as well as numerically
using the Galerkin and the Chebyshev collocation methods. The main
results are summarised and discussed in $\S$\ref{sec:sum}. To make
the paper self-contained, a brief review of analytical solutions for
the semi-infinite and channel models is included in appendix \ref{sec:appA}.

\section{\label{sec:prob}Formulation of problem}

\subsection{Basic model and governing equations}

\begin{figure}
\begin{centering}
\includegraphics[width=0.5\textwidth]{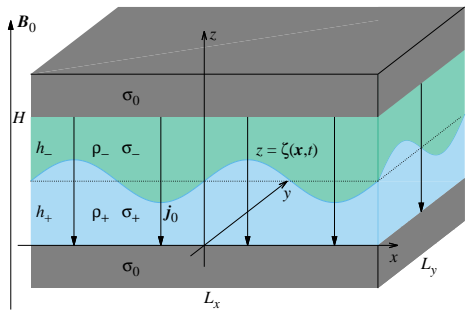}
\par\end{centering}
\caption{\label{fig:sketch}Sketch of the problem.}
\end{figure}

Consider a shallow rectangular cell of size $L_{x}\times L_{y}$ bounded
horizontally by insulating sidewalls and vertically by two solid electrodes
which are separated by a constant distance $H\ll L$, as shown in
figure \ref{fig:sketch}. The electrodes carry electric current of
a constant density $\vec{j}_{0}=-j_{0}\vec{e}_{z}$ which passes through
a layer of liquid metal (aluminium) of constant depth $\bar{h}_{+}$
and density $\rho_{+}$ superposed by a layer of electrolyte of depth
$\bar{h}_{-}=H-\bar{h}_{+}$ and density $\rho_{-}<\rho_{+}$. The
electrical conductivity of the liquid metal $\sigma_{+}$ is supposed
to be much higher than that of the electrodes $\sigma_{0}$ which,
in turn, is much higher than that of the electrolyte: $\sigma_{-}\ll\sigma_{0}\ll\sigma_{+}.$
This approximation, which is used already by \citet{Sneyd1994}, is
applicable to aluminium reduction cells because $\sigma_{-}/\sigma_{0}\sim\sigma_{0}/\sigma_{+}\sim10^{-2}$
\citep{Gerbeau2006,Molokov2011}.

The two-layer system is subject to a downward gravity force with the
free fall acceleration $g$ and external magnetic field $\vec{B}_{0}=B_{0}\vec{e}_{z}$
which is assumed to be vertical and uniform. Horizontal components
of the magnetic field, which are known to have a stabilizing effect
as long as they are predominantly generated by the current passing
through the system \citep{Sneyd1985,Herreman2019}, are as usual assumed
to be negligible for the metal pad instability. This is because the
effect of the horizontal magnetic field being determined by its horizontal
gradient \citep{Sneyd1985} is $\sim H/L=\varepsilon\ll1$ relative
to that of the vertical magnetic field which is determined by the
magnitude of this field \citep{Sneyd1994}. The system admits a stably
stratified quiescent equilibrium state with the layers separated by
a plane horizontal interface. There is no electromagnetic force in
this base state because the electric current is perfectly aligned
with the magnetic field.

Note that a magnetostatic equilibrium in this system is in general
possible if the magnetic field is vertically invariant. This, in turn,
means that the vertical magnetic field, which is supposed to be generated
by external currents, has to be also horizontally invariant \citep{Sneyd1994}.
This is not the case in the model introduced by \citet{URATA1976}
and also adopted by \citet{Bojarevics1994} who consider a horizontally
inhomogeneous vertical component of the magnetic field but still assume
a static base state.

Despite the highly idealised nature, this simple model captures the
salient features of the Sele instability mechanism which is commonly
believed to be relevant to aluminium reduction cells as well as to
large-scale liquid metal batteries. For a comprehensive discussion
of the relevance of the Sele instability mechanism (as well as a more
more detailed analysis of the assumptions underlying this model),
we refer the reader to \citet[Sec. 6.1.4.3]{Gerbeau2006} who note
the following: `Indeed, the model behind the Sele mechanism has been
successfully used to improve the stability of some industrial cells
(see {[}236{]}). In addition, the Sele mechanism has been confirmed
by complete 3-D magnetohydrodynamic (MHD) simulations (see {[}99,
101, 179{]} and Section 6.3.3).'

Let us now consider a perturbed interface with the height $z=\zeta(\vec{x},t)$
which equals the depth of the bottom layer, $h_{+}(\vec{x},t)=\zeta(\vec{x},t)$,
while the depth of the top layer is $h_{-}=H-h_{+}$. Following the
shallow-water (long-wave) approximation, we assume that the perturbation
has a characteristic longitudinal length scale $L\gg H$. Then the
mass conservation implies that the vertical velocity $w$ is of relatively
small magnitude $\sim H/L=\varepsilon\ll1$. Respectively, the associated
fluid flow is predominantly horizontal with the velocity $\vec{u}=u\vec{e}_{x}+v\vec{e}_{y}$.
Such a flow has a negligible effect on the vertical pressure distribution
which is thus purely hydrostatic:

\begin{equation}
p_{\pm}(\vec{x},z,t)=\mathit{\Pi}(\vec{x},t)+\rho_{\pm}g(\zeta-z),\label{eq:prs}
\end{equation}
where $\mathit{\Pi}(\vec{x},t)=\left.p_{\pm}(\vec{x},z,t)\right|_{z=\zeta}$
is the pressure distribution along the interface and $\vec{x}$ is
the position vector in the $(x,y)$ plane. Substituting this pressure
distribution into the horizontal fluid flow equation, we obtain 
\begin{equation}
\rho(\partial_{t}\vec{u}+\vec{u}\cdot\vec{\nabla}\vec{u})=-\vec{\nabla}(\mathit{\Pi}+\rho g\zeta)+\vec{j}\times\vec{B},\label{eq:euler}
\end{equation}
where $\vec{j}$ is the current density and the $\pm$ indices are
omitted for the sake of brevity.

The associated perturbation of the electric current density in each
layer is governed by Ohm's law for a moving medium 
\begin{equation}
\vec{j}=\sigma(\vec{E}+\vec{u}\times\vec{B}),\label{eq:ohm}
\end{equation}
where $\vec{E}$ is the perturbation of the electric field in the
laboratory frame of reference. The oscillation period of interfacial
waves, $\tau_{0}\sim L/c$, where $c$ is the wave speed (\ref{eq:c}),
is assumed to be much longer than the characteristic magnetic diffusion
time $\tau_{m}\sim\mu_{0}\sigma H^{2},$ where $\mu_{0}=\unit[4\pi\times10^{-7}]{H/m}$
is the vacuum permeability. This is the case in aluminium reduction
cells because $\tau_{m}/\tau_{0}\sim\mu_{0}\sigma cH^{2}/L\sim\varepsilon\mathit{Rm}\ll1,$
where not only $\varepsilon=H/L$ but also the magnetic Reynolds number
$\mathit{Rm=\mu_{0}\sigma cH}$ is usually small. It means that the
induction effect due to the temporal variation of the magnetic field
perturbation is negligible. Hence, the Faraday law of induction reduces
to $\vec{\nabla}\times\vec{E}=0$, which corresponds to the so-called
quasi-stationary or inductionless approximation commonly used in the
liquid-metal MHD \citep{Roberts1967}. Then we have $\vec{E}=-\vec{\nabla}\varphi$,
where $\varphi$ is the electric potential.

In addition, the perturbation of the electric field induced by the
flow across the magnetic field, $\vec{u}\times\vec{B}$, is assumed
to negligible relative to the horizontal electric field perturbation
caused by the interface deformation. Using the charge conservation
$\vec{\nabla}\cdot\vec{j}=0$, the latter can be estimated as $\sim j_{0}L/(\sigma H),$
whereas for the former we have $\sim B_{0}c,$ which thus leads to
$HB_{0}c\sigma/(Lj_{0})\ll1.$ For $B_{0}\sim\mu_{0}j_{0}L,$ this
reduces to $\mathit{Rm}\ll1,$ which, as noted above, is typically
the case, and confirms the applicability of the aforementioned inductionless
approximation. Note that in this approximation, also the magnetic
field induced by the fluid flow is negligible relative to the external
magnetic field. On the other hand, the perturbation of the vertical
magnetic field component produced by the interface deformation with
amplitude $\sim H$ can be evaluated as $\sim B_{0}H/L.$ Thus, we
have $\vec{B}=B_{0}\vec{e}_{z}+O(\varepsilon,\mathit{Rm}).$

\subsection{Reduction of the electric potential equation}

Taking into account the above estimates, Ohm's law (\ref{eq:ohm})
reduces to $\vec{j}=-\sigma\vec{\nabla}\varphi$ and, thus, the charge
conservation $\vec{\nabla}\cdot\vec{j}=0$ results in 
\begin{equation}
\vec{\nabla}^{2}\varphi=0,\label{eq:ptn}
\end{equation}
which governs the electric potential in both layers. At the interface
$z=\zeta(\vec{x},t)$, we have the continuity of both the potential
and the normal component of electric current: 
\begin{equation}
[\varphi]=[j_{n}]=0,\label{bc:ptn-i}
\end{equation}
where $[f]$ stands for a jump of $f$ across the interface. The top
electrode (anode) is assumed to be perfectly conducting relative to
the electrolyte, which means that the interface between both is effectively
equipotential. Since the electric potential is defined up an additive
constant, we can set 
\begin{equation}
\left.\varphi_{-}\right|_{z=H}=0.\label{bc:ptn-t}
\end{equation}
As the electrical conductivity of the bottom electrode (cathode) is
much lower than that of aluminium, the former may be regarded as effectively
insulating with respect to internal current perturbations in the two-layer
system. It means that the current density at the cathode is fixed
\begin{equation}
\left.\sigma_{+}\partial_{z}\varphi_{+}\right|_{z=0}=j_{0}.\label{bc:ptn-b}
\end{equation}
Since $H/L=\varepsilon\ll1$ the potential distribution in the aluminium
layer can be approximated by the power series expansion in $z$ \citep{Bojarevics1994}:
\begin{equation}
\varphi_{+}(\vec{r})=\phi_{+}^{(0)}(\vec{x})+z\phi_{+}^{(1)}(\vec{x})+\frac{z^{2}}{2}\phi_{+}^{(2)}(\vec{x})+O(\varepsilon^{3}),\label{eq:ptn-p}
\end{equation}
where $\vec{x}=\vec{r}-z\vec{e}_{z}=x\vec{e}_{x}+y\vec{e}_{y}$ and
$\phi_{+}^{(k)}(\vec{x})\equiv\left.\partial_{z}^{k}\varphi_{+}\right|_{z=0}$.
Firstly, requiring (\ref{eq:ptn-p}) to satisfy (\ref{bc:ptn-b})
and (\ref{eq:ptn}), we find 
\begin{eqnarray}
\phi_{+}^{(1)} & = & j_{0}/\sigma_{+},\label{eq:ptd1}\\
\phi_{+}^{(2)} & = & -\vec{\nabla}^{2}\phi_{+}^{(0)}+O(\varepsilon).\label{eq:ptd2}
\end{eqnarray}
Secondly, using the interface normal 
\begin{equation}
\vec{n}=\frac{\vec{e}_{z}-\vec{\nabla}h_{+}}{|\vec{e}_{z}-\vec{\nabla}h_{+}|}=\vec{e}_{z}-\vec{\nabla}h_{+}+\vec{e}_{z}O(\varepsilon^{2}),\label{eq:nz}
\end{equation}
the normal current at the interface can be expressed as 
\begin{equation}
\left.j_{n}\right|_{z=h_{+}}=-\sigma_{+}\left.\partial_{n}\varphi_{+}\right|_{z=h_{+}}\approx-\sigma_{+}(\phi_{+}^{(1)}+h_{+}\phi_{+}^{(2)}-\vec{\nabla}h_{+}\cdot\vec{\nabla}\phi_{+}^{(0)})+O(\varepsilon^{2}).\label{eq:jn}
\end{equation}
Finally, substituting (\ref{eq:ptd1}) and (\ref{eq:ptd2}) into (\ref{eq:jn}),
after a few rearrangements, we obtain

\begin{equation}
\sigma_{+}\vec{\nabla}\cdot(h_{+}\vec{\nabla}\phi_{+}^{(0)})=j_{0}+\left.j_{n}\right|_{z=h_{+}}+O(\varepsilon^{2}),\label{eq:phi0p-jn}
\end{equation}
which is a reduced two-dimensional potential equation for the aluminium
layer.

In order to determine the normal current $\left.j_{n}\right|_{z=h_{+}}$
on the RHS of (\ref{eq:phi0p-jn}), we need to consider the potential
distribution in the electrolyte layer, which can be sought as in the
aluminium layer: 
\begin{equation}
\varphi_{-}(\vec{r})=\phi_{-}^{(0)}(\vec{x})+(z-H)\phi_{-}^{(1)}(\vec{x})+\frac{1}{2}(z-H)^{2}\phi_{-}^{(2)}(\vec{x})+O(\varepsilon^{3}),\label{eq:ptn-m}
\end{equation}
where $\phi_{-}^{(k)}(\vec{x})\equiv\left.\partial_{z}^{k}\varphi_{-}\right|_{z=H}$.
Then applying (\ref{bc:ptn-t}) and (\ref{eq:ptn}), we find 
\begin{equation}
\phi_{-}^{(0)}=\phi_{-}^{(2)}=0.\label{eq:ptn02-m}
\end{equation}
Similar to the top electrode, the aluminium layer is a much better
conductor than the electrolyte layer. It means that the aluminium
is effectively equipotential with respect to the electrolyte: 
\[
\left.\varphi_{-}\right|_{z=h_{+}}=(h_{+}-H)\phi_{-}^{(1)}(\vec{x})+O(\varepsilon^{3})=\varphi_{0},
\]
where $\varphi_{0}$ is an unknown constant potential of the aluminium
layer. As a result, we have 
\[
j_{z}=-\sigma_{-}\left.\partial_{z}\varphi_{-}\right|_{z=H}=\sigma_{-}\varphi_{0}/h_{-}+O(\varepsilon^{2}),
\]
where $h_{-}=H-h_{+}$ is the depth of the electrolyte layer, and
thus 
\begin{equation}
\left.j_{n}\right|_{z=h_{+}}=j_{z}/n_{z}=j_{z}+O(\varepsilon^{2})=\sigma_{-}\varphi_{0}/h_{-}+O(\varepsilon^{2}).\label{eq:jn-h}
\end{equation}
Finally, substituting (\ref{eq:jn-h}) into (\ref{eq:phi0p-jn}),
we obtain

\begin{equation}
\sigma_{+}\vec{\nabla}\cdot(h_{+}\vec{\nabla}\phi_{+}^{(0)})=j_{0}+\sigma_{-}\varphi_{0}h_{-}^{-1},\label{eq:phi0p-0}
\end{equation}
which is accurate up to $O(\varepsilon^{2})$.

The unknown potential of the aluminium layer $\varphi_{0}$ is determined
by the solvability condition of (\ref{eq:phi0p-0}) which is required
by the Neumann boundary condition at the insulating sidewalls where
we have

\begin{equation}
\left.\partial_{n}\varphi_{+}\right|_{\Gamma}=\left.\partial_{n}\phi_{+}^{(0)}\right|_{\Gamma}+O(\varepsilon^{2})=0.\label{bc:phi0p}
\end{equation}
Integrating (\ref{eq:phi0p-0}) over the horizontal cross-section
area $S$ and using this boundary condition, we obtain

\[
\int_{S}(j_{0}+\sigma_{-}\varphi_{0}h_{-}^{-1})\thinspace\mathrm{d}^{2}\vec{x}=0,
\]
which is the solvability condition of (\ref{eq:phi0p-0}). This condition,
which requires the constancy of the total current $j_{0}S=I_{0}$,
defines the potential of the aluminium layer

\[
\varphi_{0}=-\sigma_{-}^{-1}I_{0}/\int_{S}h_{-}^{-1}\mathrm{d}^{2}\vec{x}.
\]
Substituting this into (\ref{eq:phi0p-0}), we obtain 
\begin{equation}
\sigma_{+}\vec{\nabla}\cdot(h_{+}\vec{\nabla}\phi_{+}^{(0)})=j_{0}\left(1-Sh_{-}^{-1}/\int_{S}h_{-}^{-1}\mathrm{d}^{2}\vec{x}\right),\label{eq:phi0p-hm}
\end{equation}
which is the final form of the reduced electric potential equation
for the aluminium layer with the boundary condition (\ref{bc:phi0p}).

Note that (\ref{eq:phi0p-hm}) is fully non-linear and, thus, applicable
not only to small but also to large amplitude waves. Since the present
linear stability analysis is concerned with the former, only the linearised
version of (\ref{eq:phi0p-hm}) will be required in the following.
Although the linear form of the potential equation has been derived
and used in many previous studies, the full non-linear equation (\ref{eq:phi0p-hm})
is new to the best of our knowledge. For example, \citet{Bojarevics1998}
considers non-linear waves in aluminium electrolysis cells but uses
linear potential equation (17) derived earlier by \citet{Bojarevics1994}.
Equation (\ref{eq:phi0p-hm}) differs also from analogous equation
(11) of \citet{Zikanov2000}. Instead of expanding the potential in
powers of $z$, which is essential in this case \citep{Bojarevics1994},
they integrate a three-dimensional potential equation over the depth
of the aluminium layer. As this operation cannot consistently be carried
through, \citet{Zikanov2000} effectively postulate that the integral
of the horizontal current over the depth of the aluminium layer can
be written as $\vec{J}=-\sigma_{+}\vec{\nabla}\Psi,$ whereas we have
$\vec{J}=-\sigma_{+}h_{+}\vec{\nabla}\phi_{+}^{(0)},$ which follows
from Ohm's law and the $z$-invariance of $\phi_{+}^{(0)}.$ Both
expressions are equivalent only for small-amplitude waves when $h_{+}\approx\bar{h}_{+}=\text{const}$
and, thus, $\Psi=\bar{h}_{+}\phi_{+}^{(0)}.$ Consequently, equation
(11) of \citet{Zikanov2000} is limited to small-amplitude waves.

\subsection{Linearised shallow-water model}

The distribution of the electric potential in the aluminium layer
(\ref{eq:ptn-p}) indicates that the respective electromagnetic force
\[
\vec{j_{+}}\times\vec{B_{0}}=\sigma_{+}\vec{B_{0}}\times\vec{\nabla}\phi_{+}^{(0)}(\vec{x})+O(\varepsilon^{2})
\]
is depth invariant up to $O(\varepsilon^{2})$. Since the curl of
this force has zero horizontal components, it preserves zero horizontal
vorticity of the flow. In the hydrostatic shallow-water approximation,
when the vertical velocity component is negligible, this is equivalent
to the depth invariance of the horizontal velocity: $\partial_{z}\vec{u}\equiv0$.
Consequently, this electromagnetic force is compatible with the conservation
of the depth invariance which is a generic property of the two-layer
shallow-water system \citep{Priede2020}. In the electrolyte layer,
according to (\ref{eq:ptn-m}) and (\ref{eq:ptn02-m}), the horizontal
component of the current perturbation and, thus, also the associated
electromagnetic force is $O(\varepsilon)$, i.e. negligible in the
hydrostatic shallow-water approximation. Therefore, the conservation
of the depth invariance holds also in the electrolyte layer.

The system of shallow-water equations is closed by adding the conservation
of mass in each layer: 
\begin{equation}
\partial_{t}h+\vec{\nabla}\cdot(h\vec{u})=0,\label{eq:mass}
\end{equation}
which is obtained by integrating the incompressibility constraint
$\vec{\nabla}\cdot\vec{u}=0$ over the depth of the respective layer
and using the kinematic constraint at the interface $z=\zeta(\vec{x},t):$
\[
\frac{d\zeta}{dt}=\partial_{t}\zeta+\vec{u}\cdot\vec{\nabla}\zeta=w.
\]
In the following, we focus on the interfacial waves of small amplitude:
\[
\eta(\vec{x},t)=\zeta(\vec{x},t)-\bar{h}_{+}\ll H.
\]
It means that the nonlinear terms in the governing equations are higher-order
small and, thus, negligible relative to the linear terms. After linearisation,
(\ref{eq:euler}) and (\ref{eq:mass}) take the form 
\begin{eqnarray}
\rho_{\pm}\partial_{t}\vec{u}_{\pm} & = & -\vec{\nabla}(\mathit{\Pi}+\rho_{\pm}g\eta)+\sigma_{\pm}\vec{B}_{0}\times\vec{\nabla}\phi_{\pm}^{(0)},\label{eq:flow-lin}\\
\partial_{t}\eta & = & \mp\bar{h}_{\pm}\vec{\nabla}\cdot\vec{u}_{\pm}.\label{eq:mass-lin}
\end{eqnarray}
Taking the divergence of the difference of equations (\ref{eq:flow-lin})
for the top and bottom layers, which correspond to the plus and minus
indices, and using (\ref{eq:mass-lin}), we obtain the classical wave
equation 
\begin{equation}
\partial_{t}^{2}\eta=c^{2}\vec{\nabla}^{2}\eta,\label{eq:wave}
\end{equation}
for interfacial gravity waves propagating with the speed 
\begin{equation}
c=\left(\frac{g(\rho_{+}-\rho_{-})}{\rho_{+}/\bar{h}_{+}+\rho_{-}/\bar{h}_{-}}\right)^{1/2}.\label{eq:c}
\end{equation}

Note that since the electromagnetic force is entirely solenoidal,
i.e. $\vec{B}_{0}\times\vec{\nabla}\phi_{\pm}^{(0)}=-\vec{\nabla}\times\phi_{\pm}^{(0)}\vec{B}_{0}$,
it can drive a flow without disturbing pressure as long as the flow
is not obstructed by the sidewalls. It implies that the electromagnetic
force can affect interfacial waves only at the sidewalls. The boundary
condition for (\ref{eq:wave}) follows from $\left.u_{n}\right|_{\mathit{\Gamma}}=0$
which holds at the impermeable sidewall $\mathit{\Gamma}$. Taking
again the difference of equations (\ref{eq:flow-lin}) for the top
and bottom layers to eliminate the pressure and then applying the
impermeability condition, we obtain 
\begin{equation}
\left.g[\rho]\partial_{n}\eta\right|_{\mathit{\Gamma}}=\left.B_{0}\partial_{\tau}[\sigma\phi^{(0)}]\right|_{\mathit{\Gamma}},\label{bc:eta-gen}
\end{equation}
where $\partial_{n}\equiv\vec{n}\cdot\vec{\nabla}$ and $\partial_{\tau}\equiv\vec{\tau}\cdot\vec{\nabla}$
with $\vec{n}$ and $\vec{\tau}=\vec{n}\times\vec{e}_{z}$ standing
for the outward normal and tangent vectors to the lateral cell boundary
$\mathit{\Gamma};$ the square brackets denote the difference of the
enclosed quantity between the top and bottom layers.

After linearisation, that is, applying 
\begin{eqnarray*}
h_{+} & = & \bar{h}_{+}+\eta\approx\bar{h}_{+},\\
h_{-}^{-1} & = & (\bar{h}-\eta)^{-1}\approx\bar{h}_{-}^{-1}+\eta\bar{h}_{-}^{-2},
\end{eqnarray*}
 and taking into account that $\int_{s}\eta\thinspace\mathrm{d}^{2}\vec{x}=0$
due to the mass conservation, the potential equation (\ref{eq:phi0p-hm})
reduces to

\begin{equation}
\vec{\nabla}^{2}\phi_{+}^{(0)}=-\frac{j_{0}\eta}{\sigma_{+}\bar{h}_{+}\bar{h}_{-}},\label{eq:linpot}
\end{equation}
which satisfies the solvability condition automatically owing to the
mass conservation. This equation originally formulated by \citealt{URATA1976}
differs from the analogous equation derived later by \citet{Bojarevics1994}
by the absence of terms containing the ratio of electric conductivities
of electrolyte and aluminium, $\sim\sigma_{-}/\sigma_{+}\sim10^{-4},$
which is $\sim(H/L)^{2},$ i.e. comparable to the higher-order-small
terms usually neglected in the hydrostatic shallow-water approximation
\citep{Lukyanov2001}.

Henceforth, we change to dimensionless variables by using $L=\sqrt{L_{x}L_{y}}=S^{1/2}$,
$\tau_{0}=L/c$ and $\phi_{0}=I_{0}/(\sigma_{+}\bar{h}_{+})$ as the
length, time and electric potential scales, respectively. The dimensionless
governing equations (\ref{eq:wave}) and (\ref{eq:linpot}) read as
\begin{eqnarray}
\partial_{t}^{2}\eta+\gamma\partial_{t}\eta & = & \vec{\nabla}^{2}\eta,\label{eq:eta}\\
-\eta & = & \vec{\nabla}^{2}\phi,\label{eq:phi}
\end{eqnarray}
where $\phi$ is the dimensionless counterpart of $\phi_{+}^{(0)}$.
Note that following \citet{Lukyanov2001}, we have added a linear
damping (viscous friction) term to (\ref{eq:eta}) defined by a phenomenological
friction parameter $\gamma$. The latter is subsequently assumed to
be small and, thus, ignored, unless stated otherwise. The dimensionless
boundary conditions corresponding to (\ref{bc:phi0p}) and  (\ref{bc:eta-gen})
are 
\begin{eqnarray}
\left.\partial_{n}\eta\right|_{\mathit{\Gamma}} & = & \beta\left.\partial_{\tau}\phi\right|_{\mathit{\Gamma}},\label{bc:eta}\\
\left.\partial_{n}\phi\right|_{\mathit{\Gamma}} & = & 0,\label{bc:phi}
\end{eqnarray}
where 
\begin{equation}
\beta=\frac{I_{0}B_{0}}{g(\rho_{+}-\rho_{-})\bar{h}_{+}\bar{h}_{-}}\label{def:beta}
\end{equation}
is the key dimensionless parameter for this problem, referred to by
\citet{Gerbeau2006} as the Sele number \citep{Sele1977}, which defines
the strength of electromagnetic force relative to that of gravity.

\section{\label{subsec:cell}Linear stability analysis}

\subsection{Eigenvalue perturbation solution for $\beta\ll1$}

Let us now turn to the linear stability problem posed by (\ref{eq:eta})
and (\ref{eq:phi}) for a rectangular cell of aspect ratio $\alpha=L_{x}/L_{y}$
which is bounded laterally by four sidewalls. Although the problem
does not admit an exact analytical solution as for the semi-infinite
and channel geometries (see appendix \ref{sec:appA}), it can still
be solved approximately for sufficiently small $\beta$ using the
classical eigenvalue perturbation method \citep[Sec. 1.6]{hinch_1991}.
To this end, the eigenmode 
\begin{equation}
\{\eta,\phi\}(\vec{x},t)=\{\hat{\eta},\hat{\phi}\}(\vec{x})e^{-i\omega t}+\text{c.c.},\label{eq:tmod}
\end{equation}
where c.c. stands for the complex conjugate, is sought by expanding
the eigenvalue $\omega^{2}$ and the respective amplitude distribution
in the power series of $\beta:$ 
\begin{eqnarray}
\omega^{2} & = & \lambda^{(0)}+\beta\lambda^{(1)}+\ldots,\label{eq:egv}\\
\{\hat{\eta},\hat{\phi}\}(\vec{x}) & = & \{\hat{\eta},\hat{\phi}\}^{(0)}(\vec{x})+\beta\{\hat{\eta},\hat{\phi}\}^{(1)}(\vec{x})+\ldots.\label{eq:amp}
\end{eqnarray}
At the leading order, which corresponds to $\beta=0$, (\ref{eq:eta})-{}-(\ref{bc:phi})
reduce to 
\begin{eqnarray}
\lambda^{(0)}\hat{\eta}^{(0)}+\vec{\nabla}^{2}\hat{\eta}^{(0)}=0, &  & \left.\partial_{n}\hat{\eta}^{(0)}\right|_{\mathit{\Gamma}}=0,\label{eq:eta-o0}\\
\hat{\eta}^{(0)}+\vec{\nabla}^{2}\hat{\phi}^{(0)}=0, &  & \left.\partial_{n}\hat{\phi}^{(0)}\right|_{\mathit{\Gamma}}=0.\label{eq:phi-o0}
\end{eqnarray}
The respective leading-order solution is 
\begin{eqnarray}
\lambda_{\vec{k}}^{(0)} & = & \vec{k}^{2},\label{eq:lmb0}\\
\{\hat{\eta},\hat{\phi}\}^{(0)}(\vec{x}) & = & \{\vec{k}^{2},1\}\hat{\phi}{}_{\vec{k}}^{(0)}\Psi_{\vec{k}}(\vec{x}),\label{eq:egm0}
\end{eqnarray}
where 
\begin{equation}
\Psi_{\vec{k}}(\vec{x})=\cos(xk_{x})\cos(yk_{y})\label{eq:gwmod}
\end{equation}
is the gravity wave mode with the wave vector 
\begin{equation}
\vec{k}=(k_{x},k_{y})=\pi(m/\sqrt{\alpha},n\sqrt{\alpha}),\,\,m,n=0,1,2,\ldots\label{eq:wavec}
\end{equation}

It is important to note that some leading-order eigenmodes (\ref{eq:egm0})
may be degenerate, namely, to have the same eigenfrequency (\ref{eq:lmb0})
for different wave vectors. In this case, the leading-order solution
can be a superposition of the eigenmodes (\ref{eq:egm0}) with the
same magnitude of the wave vector $\vec{k}'^{2}=\vec{k}^{2}:$ 
\begin{equation}
\{\hat{\eta},\hat{\phi}\}^{(0)}(\vec{x})={\displaystyle \sum_{\vec{\vec{k}'^{2}=\vec{k}^{2}}}\{\hat{\eta},\hat{\phi}\}_{\vec{k'}}^{(0)}\Psi_{\vec{k'}}(\vec{x}).}\label{eq:egs0}
\end{equation}

Following the standard approach, the first-order correction $\{\hat{\eta},\hat{\phi}\}^{(1)}$
to (\ref{eq:egs0}) is sought as an expansion in the leading-order
eigenmodes 
\[
\{\hat{\eta},\hat{\phi}\}^{(1)}(\vec{x})={\displaystyle \sum_{\vec{k}}\{\hat{\eta},\hat{\phi}\}_{\vec{k}}^{(1)}\Psi_{\vec{k}}(\vec{x}),}
\]
where the summation is over all wave vectors. Substituting this expansion
into the first-order problem
\begin{eqnarray}
\lambda^{(0)}\hat{\eta}^{(1)}+\vec{\nabla}^{2}\hat{\eta}^{(1)}=-\lambda^{(1)}\hat{\eta}^{(0)}, &  & \left.\partial_{n}\hat{\eta}^{(1)}\right|_{\mathit{\Gamma}}=\left.\partial_{\tau}\hat{\phi}^{(0)}\right|_{\mathit{\Gamma}},\label{eq:eta-o1}\\
\hat{\eta}^{(1)}+\vec{\nabla}^{2}\hat{\phi}^{(1)}=0, &  & \left.\partial_{n}\hat{\phi}^{(1)}\right|_{\mathit{\Gamma}}=0.\label{eq:phi-o1}
\end{eqnarray}
and using the orthogonality of leading-order eigenmodes by projecting
the result onto $\Psi_{\vec{k}}$, after a few rearrangements, we
obtain the solvability condition of the first-order problem
\begin{equation}
{\displaystyle \lambda_{\vec{k}}^{(1)}\vec{k}^{2}\left\langle \Psi_{\vec{k}}^{2}\right\rangle \hat{\phi}_{\vec{k}}^{(0)}-\sum_{\vec{k}'^{2}=\vec{k}^{2}}}F_{\vec{k},\vec{k}'}\hat{\phi}_{\vec{k}'}^{(0)}=(\vec{k}^{2}-\lambda_{\vec{k}}^{(0)})\vec{k}^{2}\left\langle \Psi_{\vec{k}}^{2}\right\rangle \hat{\phi}_{\vec{k}}^{(1)}=0,\label{eq:lmb1-gen}
\end{equation}
where the right-hand side is zero owing to the leading-order solution
(\ref{eq:lmb0}) and, thus, the left-hand side defines the eigenvalue
perturbation $\lambda_{\vec{k}}^{(1)}.$ The summation on the left-hand
side is over the degenerate modes as in (\ref{eq:egs0}); for non-degenerate
modes, it reduces to a single term with $\vec{k}'=\vec{k}$. The angle
brackets denote integral over $S=L_{x}\times L_{y}:$ 
\[
\left\langle \Psi_{\vec{k}}^{2}\right\rangle ={\displaystyle \int_{S}\Psi_{\vec{k}}^{2}\thinspace\mathrm{d}^{2}\vec{x}}=c_{k_{x}}^{-1}c_{k_{y}}^{-1},
\]
where $c_{0}=1$ and $c_{k}=2$ for $k\not=0$. The second left-hand
side term of (\ref{eq:lmb1-gen}) is due to Green's first identity
\[
\left\langle \Psi_{\vec{k}}\vec{\nabla}^{2}\hat{\eta}^{(1)}+\vec{\nabla}\Psi_{\vec{k}}\cdot\vec{\nabla}\hat{\eta}^{(1)}\right\rangle =\ointop_{\mathit{\Gamma}}\Psi_{\vec{k}}\partial_{n}\hat{\eta}^{(1)}\mathrm{d}\mathit{\Gamma},
\]
where the boundary integral can be transformed using the boundary
condition (\ref{eq:eta-o1}) and Green's theorem as follows: 
\[
\ointop_{\mathit{\Gamma}}\Psi_{\vec{k}}\partial_{\tau}\hat{\phi}^{(0)}\mathrm{d}\mathit{\Gamma}=-\left\langle \vec{e}_{z}\cdot\vec{\nabla}\Psi_{\vec{k}}\times\vec{\nabla}\hat{\phi}^{(0)}\right\rangle .
\]
This results in the electromagnetic interaction matrix 
\begin{equation}
F_{\vec{k},\vec{k}'}=\left\langle \vec{e}_{z}\cdot\vec{\nabla}\Psi_{\vec{k}}\times\vec{\nabla}\Psi_{\vec{k}'}\right\rangle =G_{\vec{k},\vec{k}'}-G_{\vec{k}',\vec{k}},\label{eq:fmtr}
\end{equation}
where $G_{\vec{k},\vec{k}'}=H_{k_{x},k_{x}'}H_{k_{y}',k_{y}}$ and
\begin{equation}
H_{k_{m},k_{n}}=k_{m}\left\langle \sin(xk_{m})\cos(xk_{n})\right\rangle =\frac{2m^{2}}{m^{2}-n^{2}}\text{mod}(m+n,2),\label{eq:hmtr}
\end{equation}
Expressions (\ref{eq:fmtr}) and (\ref{eq:hmtr}) differ form of those
used by \citet{Sneyd1994} only by a more compact form.

Since the electromagnetic interaction matrix (\ref{eq:fmtr}) is anti-symmetric,
as it is well known, there is no electromagnetic back-reaction on
separate gravity wave modes, i.e. $F_{\vec{k},\vec{k}}=0$. Thus,
for a non-degenerate (single) mode, we have 
\[
\lambda_{\vec{k}}^{(1)}\left\langle \Psi_{\vec{k}}^{2}\right\rangle k^{2}\hat{\phi}_{\vec{k}}=F_{\vec{k},\vec{k}}\hat{\phi}_{\vec{k}}=0,
\]
which means no electromagnetic effect at the first order in $\beta$.
For a degenerate mode consisting of a superposition of two eigenmodes
with the same frequency 
\begin{equation}
\lambda_{\vec{k}}^{(0)}=\vec{k}_{1}^{2}=\vec{k}_{2}^{2},\label{eq:degen}
\end{equation}
equation (\ref{eq:lmb1-gen}) can be written in the matrix form as
\begin{equation}
\lambda_{\vec{k}}^{(1)}k^{2}\left(\begin{array}{c}
\left\langle \Psi_{\vec{k}_{1}}^{2}\right\rangle \hat{\phi}_{\vec{k_{1}}}\\
\left\langle \Psi_{\vec{k}_{2}}^{2}\right\rangle \hat{\phi}_{\vec{k}_{2}}
\end{array}\right)=\left(\begin{array}{cc}
0 & F_{\vec{k}_{1},\vec{k}_{2}}\\
-F_{\vec{k}_{1},\vec{k}_{2}} & 0
\end{array}\right)\left(\begin{array}{c}
\hat{\phi}_{\vec{k}_{1}}\\
\hat{\phi}_{\vec{k}_{2}}
\end{array}\right).\label{eq:lmb1-2}
\end{equation}
The solution of this second-order matrix eigenvalue problem yields
\begin{equation}
\lambda_{\vec{k}}^{(1)}k^{2}\sqrt{\left\langle \Psi_{\vec{k}_{1}}^{2}\right\rangle \left\langle \Psi_{\vec{k}_{2}}^{2}\right\rangle }=\pm iF_{\vec{k}_{1},\vec{k}_{2}},\label{eq:lmb1}
\end{equation}
which indicates that $\lambda_{\vec{k}}^{(1)}$ is imaginary and,
thus, the system is unstable provided that $F_{\vec{k_{1}},\vec{k}_{2}}\not=0$.
According to (\ref{eq:hmtr}), this is the case only if both components
of the two wave vectors $\vec{k}_{1`}=\pi(m_{1}/\sqrt{\alpha},n_{1}\sqrt{\alpha})$
and $\vec{k}_{2}=\pi(m_{2}/\sqrt{\alpha},n_{2}\sqrt{\alpha})$ have
opposite parities; namely, $m_{1}\pm m_{2}$ and $n_{1}\pm n_{2}$
are odd numbers. Then the degeneracy condition (\ref{eq:degen}) yields
\[
\alpha_{c}^{2}=-\frac{(m_{1}+m_{2})(m_{1}-m_{2})}{(n_{1}+n_{2})(n_{1}-n_{2})}=\frac{m}{n},
\]
where $m$ and $n$ can be any odd numbers. Consequently, all cells
with aspect ratios squared equal to the ratio of two odd numbers are
inherently unstable, i.e. they become unstable at infinitesimal $\beta>\beta_{c}=0$.

For $\alpha^{2}=1$, which corresponds to a square cell, the unstable
wavenumbers are $m_{1}=l,\,n_{1}=0$ and $m_{2}=0,\,n_{2}=l$, where
$l=1,3,5,\ldots$. In this case, (\ref{eq:lmb1}) yields $\lambda_{l}^{(1)}=\pm i8/(l\pi)^{2}$
and hence the instability growth rate resulting from (\ref{eq:egv})
is 
\begin{equation}
\Im[\omega_{l}^{(1)}]=\pm\frac{4}{(l\pi)^{3}},\label{eq:grt1-l}
\end{equation}
which means that the most unstable is the mode with $l=1$. For a
general $\alpha^{2}=m/n$ defined by arbitrary odd numbers $m$ and
$n$, the lowest unstable wavenumbers can be written as
\begin{eqnarray}
m_{(3\pm1)/2} & = & (m\pm1)/2,\label{eq:m12}\\
n_{(3\pm1)/2} & = & (n\mp1)/2.\label{eq:n12}
\end{eqnarray}
In this case, (\ref{eq:lmb1}) yields $\lambda_{m,n}^{(1)}=\pm i8c_{m-1}^{1/2}c_{n-1}^{1/2}/(\pi^{2}\sqrt{mn})$
and, respectively, (\ref{eq:egv}) results in 
\begin{equation}
\Im[\omega_{m,n}^{(1)}]=\pm\frac{8}{\pi^{3}}\left(\frac{c_{m-1}c_{n-1}}{(m+n)(1+mn)\sqrt{mn}}\right)^{1/2},\label{eq:grt1-a}
\end{equation}
which reduces to (\ref{eq:grt1-l}) with $l=1$ when $m=n=1$. It
is important to note that the growth rate (\ref{eq:grt1-a}) drops
off with the increase of $m$ and $n$ as $\sim(m+n)^{-1/2}(mn)^{-3/4}.$

For an aspect ratio $\alpha$ sufficiently close to the critical value
$\alpha_{c}=\sqrt{m/n},$ the stability of system is expected to be
determined by the interaction of two critical modes with the wavenumbers
$($\ref{eq:m12},\ref{eq:n12}) which correspond to the wavenumbers
\[
\vec{k}_{(3\pm1)/2}=\frac{\pi}{2}((m\pm1)/\sqrt{\alpha},(n\mp1)\sqrt{\alpha}).
\]
Searching the amplitude distribution in (\ref{eq:tmod}) as 
\[
\{\hat{\eta},\hat{\phi}\}(\vec{x})=\{\hat{\eta},\hat{\phi}\}_{\vec{k}_{1}}\Psi_{\vec{k}_{1}}+\{\hat{\eta},\hat{\phi}\}_{\vec{k}_{2}}\Psi_{\vec{k}_{2}},
\]
without assuming $\beta$ to be small, we obtain the following second-order
matrix eigenvalue problem for $\omega^{2}:$ 
\begin{equation}
\left(\begin{array}{cc}
(\omega^{2}-\vec{k}_{1}^{2})\vec{k}_{1}^{2} & \beta F_{\vec{k}_{1},\vec{k}_{2}}c_{n-1}\\
-\beta F_{\vec{k}_{1},\vec{k}_{2}}c_{m-1} & (\omega^{2}-\vec{k}_{2}^{2})\vec{k}_{2}^{2}
\end{array}\right)\left(\begin{array}{c}
\hat{\phi}_{\vec{k}_{1}}\\
\hat{\phi}_{\vec{k}_{2}}
\end{array}\right)=\mathsf{\mathbf{0}},\label{eq:eigv2}
\end{equation}
where $F_{\vec{k}_{1},\vec{k}_{2}}=2(m+n)(mn+1)/(mn)$. As before,
for the system to be stable, the eigenvalue $\omega^{2}$ has to be
real. This is the case if $\beta\le\beta_{m,n}(\alpha^{2})$, where
\begin{align}
\beta_{m,n}(\alpha^{2})= & \frac{\pi^{4}}{16}\frac{\left|\alpha_{c}^{2}/\alpha{}^{2}-1\right|mn^{2}}{(m+n)(mn+1)}\nonumber \\
\times & \left(\frac{(m^{2}-1)^{2}+\alpha^{4}(n^{2}-1)^{2}+2\alpha^{2}((m+n)^{2}+(mn+1)^{2})}{c_{m-1}c_{n-1}}\right)^{1/2}.\label{eq:bet-mn}
\end{align}
For aspect ratios with $\alpha^{2}=m$, where $m$ is an even number,
using the two-mode approximation (\ref{eq:bet-mn}) for the preceding
major critical point, i.e. $\alpha_{c}^{2}=m-1$, we have 
\[
\beta_{m-1,1}(m)=\frac{\pi^{4}}{16}\frac{(m-1)}{m^{2}}\sqrt{\frac{m^{2}+4}{c_{m-2}}}.
\]
The lowest $\beta_{m-1,1}=\frac{3\pi^{4}}{128}\sqrt{\frac{5}{2}}\approx3.61$
occurs at $m=4$, whereas the highest $\beta_{m-1,1}=\frac{\pi^{4}}{16\sqrt{2}}\approx4.30$
at $m=2$, which is then approached asymptotically at large even $m$.

The intersection of instability thresholds associated with the first
two major critical points, which is defined by $\beta_{1,1}(\alpha^{2})=\beta_{3,1}(\alpha^{2})$,
yields $\alpha_{c}^{2}\approx2.125$ and, thus, in the two-mode approximation,
we have $\beta_{c}\approx4.699$. As shown in the next section, this
is the optimal aspect ratio and the corresponding highest stability
threshold which can be achieved with a small viscous friction.

\subsection{Numerical solution of the matrix eigenvalue problem}

\begin{figure}
\begin{centering}
\includegraphics[width=0.5\columnwidth]{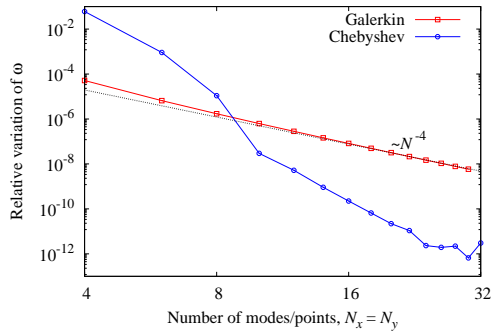}
\par\end{centering}
\caption{\label{fig:err}The relative variation of the complex frequency with
the largest imaginary part ($\omega=3.14455+\mathrm{i}0.12889)$ versus
the number of modes and nodes used in the Galerkin and Chebyshev approximations
for $\alpha=\beta=1$.}
\end{figure}

For arbitrary $\beta$, (\ref{eq:tmod}) leads to 
\begin{eqnarray}
\lambda\hat{\eta}+\vec{\nabla}^{2}\hat{\eta}=0, &  & \left.\partial_{n}\hat{\eta}\right|_{\mathit{\Gamma}}=\beta\left.\partial_{\tau}\hat{\phi}\right|_{\mathit{\Gamma}},\label{eq:etah2}\\
\hat{\eta}+\vec{\nabla}^{2}\hat{\phi}=0, &  & \left.\partial_{n}\hat{\phi}\right|_{\mathit{\Gamma}}=0,\label{eq:phih2}
\end{eqnarray}
which is an eigenvalue problem for $\lambda=\omega^{2}+\mathrm{i}\gamma$,
where $\gamma$ is a viscous friction coefficient. The problem can
be discretised using the Galerkin method with the gravity wave modes
(\ref{eq:gwmod}) as basis functions. This leads to the generalisation
of (\ref{eq:lmb1-gen}): 
\[
(\lambda-\vec{k}^{2})\vec{k}^{2}\left\langle \Psi_{\vec{k}}^{2}\right\rangle \hat{\phi}_{\vec{k}}=\beta{\displaystyle \sum_{\vec{k}'}}F_{\vec{k},\vec{k}'}\hat{\phi}_{\vec{k}'},
\]
with the electromagnetic interaction matrix on the right-hand side
defined by (\ref{eq:fmtr}). This is a matrix eigenvalue problem of
size $(M+1)^{2}\times(N+1)^{2}$, where $M$ and $N$ are the cutoff
limits of the $x$- and $y$-components of the wave vectors (\ref{eq:wavec}).
This method with $M\times N=12$ modes was used by \citet{Sneyd1994}
to compute the stability threshold for a cell of length $\unit[7.7]{m}$
depending on the width which varied from $\unit[\approx0.5]{m}$ to
$\unit[8]{m}.$ Alternatively, (\ref{eq:etah2},\ref{eq:phih2}) can
be discretised using the Chebyshev collocation method \citep{Boyd2013}
(see appendix \ref{sec:appB}). As seen in figure \ref{fig:err},
the latter has a significantly faster convergence rate than $\sim N^{-4}$
achieved by the Galerkin approximation with $N$ modes in each direction.
As the relative accuracy of the Chebyshev collocation approximation
saturates at $\approx10^{-11}\cdots10^{-12}$ when $N\apprge24$,
in the following, we use $16\cdots24$ collocation points in each
direction.

\begin{figure}
\begin{centering}
\includegraphics[width=0.5\columnwidth]{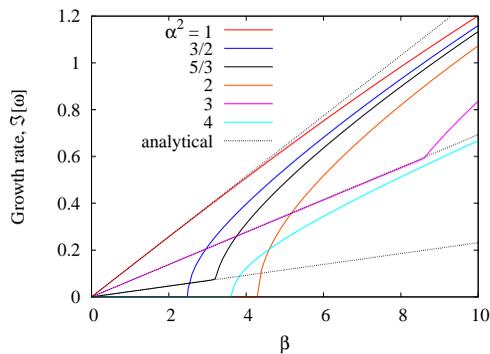}
\par\end{centering}
\caption{\label{fig:grt-w}The highest growth rate $\Im[\omega]$ depending
on the interaction parameter $\beta$ for various aspect ratios $\alpha$
computed by the Chebyshev collocation method with $N_{x}=N_{y}=16$.
For $\alpha^{2}=m/n$, where $m$ and $n$ are odd numbers, numerical
results are compared with the approximate analytical solution (\ref{eq:grt1-a}).}
\end{figure}

\begin{figure}
\begin{centering}
\includegraphics[width=0.75\columnwidth]{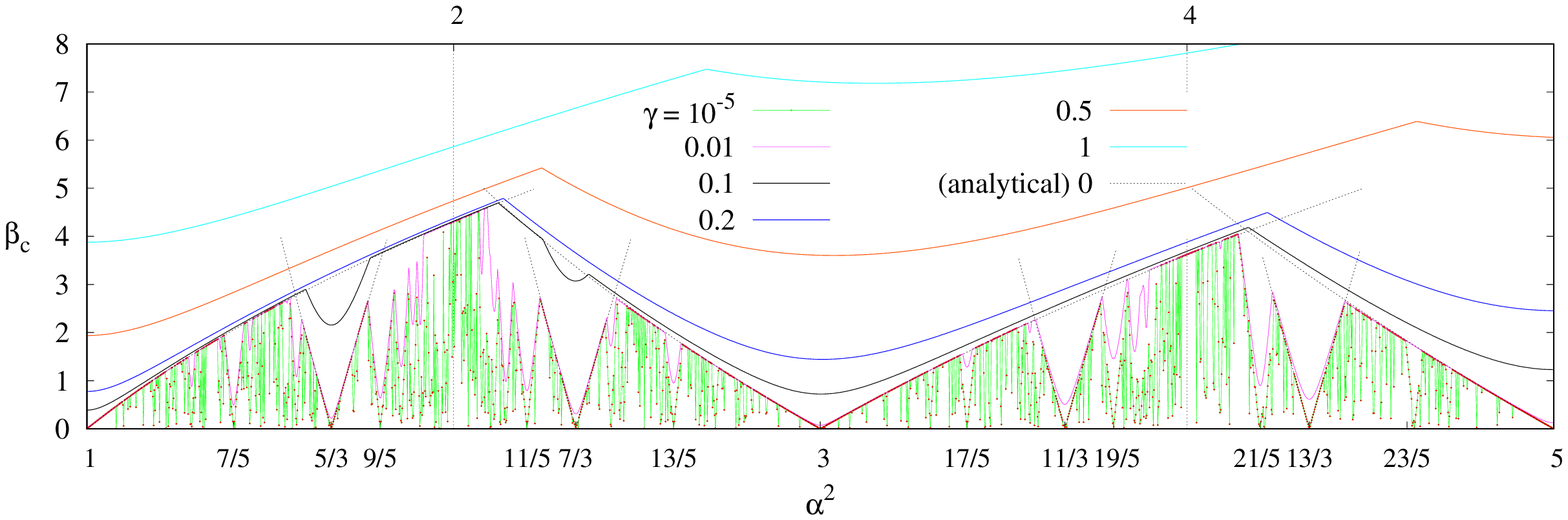}
\par\end{centering}
\centering{}\includegraphics[width=0.75\columnwidth]{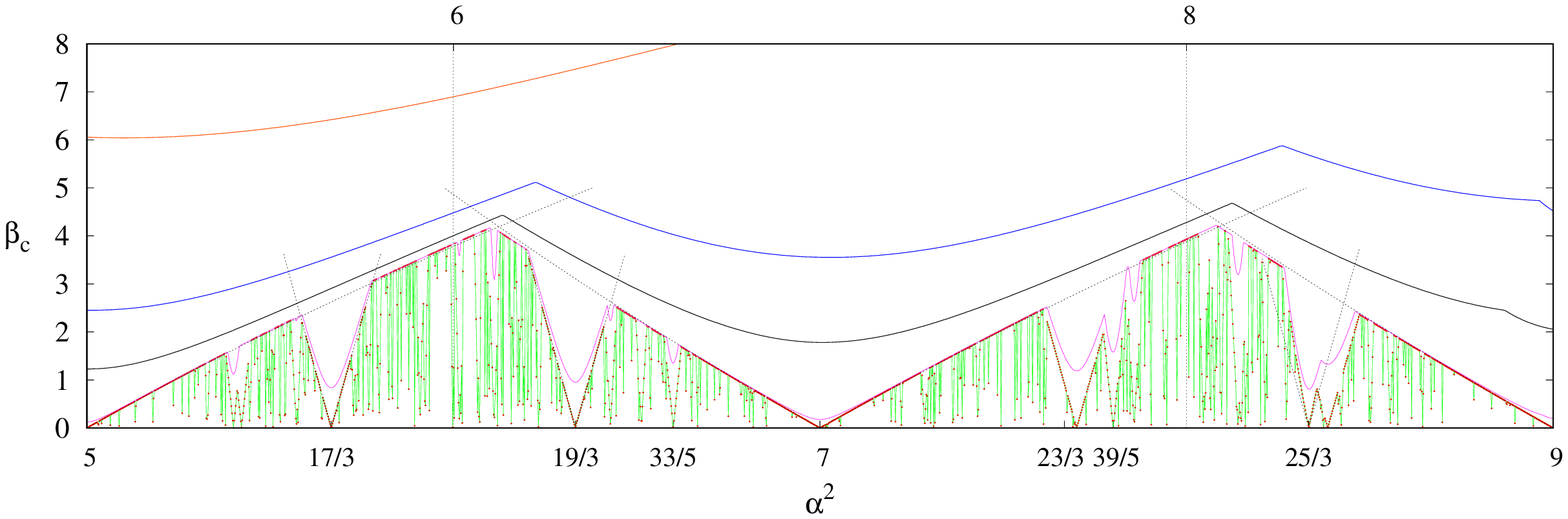} \caption{\label{fig:alphabet-c}Instability threshold $\beta_{c}$ depending
on the aspect ratio squared $(\alpha^{2})$ computed for various viscous
friction coefficients $\gamma$ using Chebyshev collocation method
with $N_{x}=N_{y}=16\cdots24$ points. Analytical solution (\ref{eq:bet-mn})
is plotted for the dominant critical points $\alpha_{c}^{2}$ equal
to odd numbers and their thirds.}
\end{figure}
The highest growth rate $\omega_{i}=\Im[\omega]$, which is computed
using Chebyshev collocation method with $N_{x}=N_{y}=16$ points and
plotted in figure \ref{fig:grt-w} against the interaction parameter
$\beta$ for various aspect ratios $\alpha$, confirms the eigenvalue
perturbation solution obtained in the previous section. Namely, for
$\alpha^{2}$ equal to the ratio of two odd numbers, the growth rate
becomes positive at $\beta>\beta_{c}=0$ whereas for other aspect
ratios this happens only at finite $\beta_{c}$. The dependence of
the instability threshold $\beta_{c}$ on $\alpha^{2}$ is shown in
figure \ref{fig:alphabet-c} for various viscous friction coefficients
$\gamma$. Note that the instability threshold cannot be computed
for $\gamma=0$ because it becomes absolutely fractal in this limit.
Due to the finite accuracy of numerical solution, the lowest friction
coefficient for which the stability threshold can reliably be computed
is around $\gamma=10^{-5}.$ For this value, the stability diagram
is very rugged and exhibits both small- and large-scale patterns.
The key feature are the dips in $\beta_{c}$ which can be seen to
occur at $\alpha_{c}^{2}$ equal to the ratio of two odd numbers as
predicted by the eigenvalue perturbation analysis in the previous
section. The increase of the friction coefficient $\gamma$ gradually
smooths out the dependence of $\beta_{c}$ on $\alpha$, especially
at small scales and larger aspect ratios. However, the main feature
of the stability diagram, which is the location of minima and maxima
of $\beta_{c}$ in the vicinity of odd and even values of $\alpha^{2}$,
respectively, persists up to relatively large friction coefficients
$\gamma\sim0.1$. It is remarkable that the approximate solution (\ref{eq:bet-mn})
for the major critical points closely reproduces the numerical results
obtained with $N_{x}=N_{y}=24$ collocation points. According to the
approximate solution obtained in the previous section, the first four
maxima of $\beta_{c}\approx4.70,4.11,4.19,4.23$ are located at $\alpha^{2}\approx2.13,4.16,6.12,8.10$.

\section{\label{sec:sum}Summary and discussion}

We have carried out a comprehensive linear stability analysis of interfacial
waves in the idealised model of an aluminium reduction cell. The model
consists of two stably stratified liquid layers carrying a vertical
electric current in the presence of a uniform collinear external magnetic
field. Using the long-wave approximation, we derived a fully nonlinear
shallow-water model comprising a novel integro-differential equation
for the electric potential. Our shallow-water model differs from that
of \citet{Bojarevics1998} by being fully nonlinear and, thus, applicable
not only to small but also to large amplitude waves. It should also
be noted that the electric potential equation we derived differs from
that postulated by \citet{Zikanov2000}. The nonlinear potential equation
is just a side result which may be useful for numerical modelling
of large amplitude shallow-water waves but it is not essential for
the linear stability analysis of infinitesimal amplitude waves considered
in the present study.

In this study, we showed that rectangular cells are inherently unstable
if their aspect ratio squared equals the ratio of two odd numbers:
$\alpha^{2}=m/n.$ In the inviscid limit, such cells can be destabilised
by an infinitesimally weak electromagnetic interaction while cells
with other aspect ratios have finite instability thresholds. The
unstable aspect ratios form an absolutely discontinuous dense set
of points which intersperse aspect ratios with finite stability thresholds.
This inherent fractality of the instability threshold was revealed
analytically using an eigenvalue perturbation method and confirmed
by accurate numerical solution of the underlying linear stability
problem.

Although the instability threshold is gradually smoothed out by viscous
friction, its principal structure, which is dominated by major critical
aspect ratios corresponding to moderate values of $m$ and $n$, is
well preserved up to relatively large dimensionless linear damping
coefficients $\gamma\sim0.1$. For $\gamma\lesssim1,$ cells with
such aspect ratios have the lowest stability threshold with the critical
electromagnetic parameter $\beta_{c}\sim\gamma,$ while the most stable
are cells with $\alpha^{2}\approx2.13$ which have $\beta_{c}\approx4.7.$

For a reduction cell operated at a total electric current of $I_{0}\approx\unit[300]{kA}$
with the electrolyte and aluminium layers of depth $\bar{h}_{-}\approx\unit[5]{cm}$
and $\bar{h}_{+}\approx\unit[25]{cm}$, respectively, and the density
difference $\Delta\rho=\rho_{+}-\rho_{-}\approx\unit[200]{kg/m^{3},}$
this $\beta_{c}$ corresponds to the critical magnetic field 
\[
B_{0}=\beta_{c}g\Delta\rho\bar{h}_{+}\bar{h}_{-}/I_{0}\approx\unit[0.27]{mT}.
\]
This field is an order of magnitude stronger than Earth's magnetic
field but still an order of magnitude weaker than the magnetic fields
which can be generated by the electric current flowing in the reduction
cell itself and in the adjacent cells. It should be noted that it
is not the whole magnetic field but only its vertical component which
is important for the metal pad instability.

It is also important to note that the above result is based on the
assumption of a relatively small linear damping in the two-layer system,
i.e. $\gamma\sim\tau_{0}/\tau_{\nu}\lesssim0.1$, where $\tau_{0}$
and $\tau_{\nu}$ are the characteristic wave propagation and viscous
damping times with the latter being the reciprocal of the linear friction
coefficient. For a cell with $L_{x}L_{y}\approx\unit[40]{m^{2}}$
and $c\approx\unit[0.2]{m/s,}$ we have $\tau_{0}=L/c\sim\unit[10]{s.}$
Viscous damping is expected to be dominated by the electrolyte layer,
which is thinner and also more viscous than the aluminium layer. Using
$\nu_{-}\sim\unit[10^{-6}]{m^{2}/s}$ as a characteristic kinematic
viscosity of cryolite \citep{Hertzberg2013}, we obtain $\tau_{\nu}\sim\bar{h}_{-}^{2}/\nu_{-}\sim\unit[10^{3}]{s}$.
On the other hand, assuming $B_{0}\sim\unit[1]{mT,}$we obtain the
same order-of-magnitude magnetic damping time in the aluminium layer:
$\tau_{m}\sim\rho_{+}/\sigma_{+}B_{0}^{2}\sim\unit[10^{3}]{s}$. Consequently,
we have $\gamma\sim10^{-2}$ for both the viscous and the magnetic
damping. This, however, is at odds with the customary assumption that
the viscous damping time in real reduction cells is $\tau_{\nu}\sim\unit[10]{s}$
and, thus, $\gamma\sim1$ \citep{Moreau1988,Zikanov2000,Molokov2011}.

Such a high damping may be due to turbulence which is likely to develop
when the waves become sufficiently large. Damping can also be affected
by the large-scale circulation which occurs in real cells due to the
non-uniformity of electric current and magnetic field. The same non-uniformity
can also affect the instability threshold directly by modifying the
electromagnetic coupling strength of different gravity wave modes.
These, however, are rather complex effects which are outside the scope
of the basic Sele instability model. Nevertheless, this model is still
relevant to the stability of real aluminium reduction cells as confirmed
by complete 3-D MHD simulations \citep[Sec. 6.1.4.3 and references therein]{Gerbeau2006}.

When the background circulation is negligible, which is usually assumed
to be the case in this model, the waves can have a sufficiently small
initial amplitude for the linear viscous friction to be relevant.
In the idealised model with uniform vertical current and a strictly
collinear background magnetic adopted in this study, turbulent damping
is a nonlinear effect which can be modelled with a velocity-dependent
turbulent friction coefficient \citep{Bojarevics2006}. Although this
effect can potentially halt the growth of instability once the waves
reach a certain amplitude, it is not expected to affect the onset
of instability. Moreover, if the saturation amplitude is relatively
small, the waves may remain practically unnoticeable well above the
linear stability threshold. Nonlinear evolution and, especially, the
saturation of metal pad instability are open questions which can be
addressed using the fully nonlinear shallow-water model proposed in
this paper. This, however, is outside the scope of the present paper
which is focused on the linear stability of two-layer system. It is
also relatively straightforward to extend this study to three-layer
systems like liquid metal batteries, which will be considered elsewhere.

\begin{acknowledgements}
\textbf{Declaration of Interests.} The authors report no conflict of interest.
\end{acknowledgements} 

\appendix

\section{\label{sec:appA}Linear stability of semi-infinite and channel geometries}

Since the electromagnetic interaction parameter appears only in the
boundary condition (\ref{bc:eta}), the destabilisation of small-amplitude
interfacial waves in the presence of a uniform vertical magnetic field
is essentially a boundary effect \citep{Lukyanov2001}. It means that
lateral boundaries are critical for the metal pad instability. As
pointed out by \citet{Lukyanov2001}, this fact is obscured by the
weak mathematical formulation of the problem where boundary conditions
are represented by integral volumetric terms \citep{Bojarevics1994}.
It is also missed by the electromechanical pendulum models where the
liquid metal layer is substituted by a solid slab \citep{Davidson1998,Zikanov2015}.
The simplest hydrodynamic model of the metal pad instability is provided
by a semi-infinite system $(0<y<\infty)\times(-\infty<x<\infty)$
bounded by a single lateral wall at $y=0$.

Let us briefly review this elementary model which was first considered
by \citet{Lukyanov2001} and later revisited by \citet{Molokov2011}.
Owing to the $x$-invariance and stationarity of the base state, a
small-amplitude disturbance of the interface height $\eta$ and the
associated electric potential $\phi$ can be sought as the normal
mode 
\begin{equation}
\{\eta,\phi\}(\vec{x},t)=\{\hat{\eta},\hat{\phi}\}(y)e^{i(kx-\omega t)}\label{eq:nmod}
\end{equation}
where $k$ is a real wavenumber describing harmonic variation of coupled
interface and potential perturbation in the longitudinal $(x)$ direction,
$\omega$ is a generally complex frequency, and $\hat{\eta}(y)$ and
$\hat{\varphi}(y)$ are amplitude distributions in the transverse
$(y)$ direction. Then (\ref{eq:eta}) and (\ref{eq:phi}) for the
perturbation amplitudes take the form 
\begin{eqnarray}
-\omega^{2}\hat{\eta} & = & \hat{\eta}''-k^{2}\hat{\eta},\label{eq:etah}\\
-\hat{\eta} & = & \hat{\phi}''-k^{2}\hat{\phi},\label{eq:phih}
\end{eqnarray}
while the boundary conditions at $y=0$ read as 
\begin{equation}
\hat{\eta}'(y)-ik\beta\hat{\phi}(y)=\hat{\phi}'(y)=0.\label{bc:etah}
\end{equation}
The solution of (\ref{eq:etah}) and (\ref{eq:phih}) bounded at $y\rightarrow\infty$
can be written in the general form as 
\begin{eqnarray}
\hat{\eta}(y) & = & \hat{\eta}_{-}e^{-i\kappa y}+\hat{\eta}_{+}e^{i\kappa y},\label{gs:etah1}\\
\hat{\phi}(y) & = & \hat{\phi}_{0}e^{-ky}+\omega^{-2}\hat{\eta}(y),\label{gs:phih1}
\end{eqnarray}
where $\hat{\eta}_{\pm}$ and $\hat{\phi}_{0}$ are unknown constants
and $\kappa=\sqrt{\omega^{2}-k^{2}}$ is the wavenumber of perturbation
in the transverse $(y)$ direction. There are two types of solutions
possible. The first corresponds to a real $\kappa$ and describes
pure gravity waves with real frequency $\omega=\pm\sqrt{k^{2}+\kappa^{2}}$.
For this solution, boundary conditions (\ref{bc:etah}) yield 
\begin{equation}
\frac{\hat{\eta}_{+}-\hat{\eta}_{-}}{\hat{\eta}_{+}+\hat{\eta}_{-}}=\frac{k\beta}{\kappa(\omega^{2}-i\beta)},\label{eq:rflx}
\end{equation}
which defines the ratio of amplitudes forming the eigenmode (\ref{eq:nmod}).
As seen, the electromagnetic effect breaks the reflection symmetry,
$\hat{\eta}_{+}=\hat{\eta}_{-}$, which holds in the non-magnetic
case $(\beta=0)$. With $\beta\not=0,$ the amplitude and phase of
the reflected wave are no longer the same as those of the incident
wave. However, the amplification of the reflected wave, which is possible
in this case, does not mean instability as suggested by \citet{Lukyanov2001}.
The stability of the system is defined by the frequency of the eigenmode
rather than by its amplitude distribution. Namely, the eigenmode is
neutrally stable as long as the frequency is real. Also, note that
for real frequencies, there is no physical distinction between the
incident and reflected waves which can be swapped one with the other
because of the time inversion symmetry. Therefore the amplification
of the reflected wave is not the mechanism behind the metal pad instability.

\begin{figure}
\begin{centering}
\includegraphics[width=0.5\columnwidth]{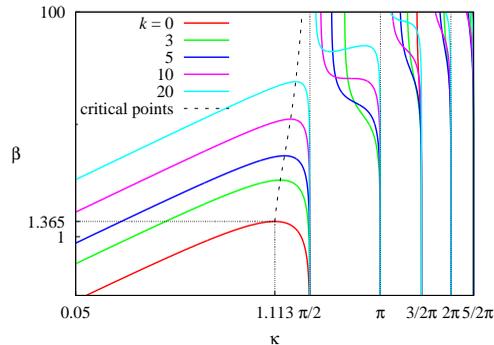}
\par\end{centering}
\caption{\label{fig:kappa}The electromagnetic interaction parameter $\beta$
versus the transverse wavenumber $\kappa$ for various longitudinal
wavenumbers $k$.}
\end{figure}

There is another solution branch which explicitly describes interfacial
wave instability. This genuinely unstable mode, which was missed by
\citet{Lukyanov2001} and considered first by \citet{Morris2003},
is defined by the complex frequencies $\omega$. In this case, the
general solution of (\ref{eq:etah}) decaying at $y\rightarrow\infty$
can be written as 
\begin{equation}
\hat{\eta}(y)=\hat{\eta}_{0}e^{-y\kappa},\label{gs:etah1i}
\end{equation}
where $\kappa=\sqrt{k^{2}-\omega^{2}}$ represents the branch with
positive real part. Substituting the solution above along with (\ref{gs:phih1})
into the boundary conditions (\ref{bc:etah}), after a few rearrangements,
we obtain the dispersion relation $\kappa(\kappa+k)+i\beta=0$, which
yields 
\[
\kappa=\sqrt{\frac{k^{2}}{4}-i\beta}-\frac{k}{2}.
\]
The respective frequency $\omega=\pm\sqrt{k^{2}-\kappa^{2}}$ then
follows from the definition of $\kappa$ given below (\ref{gs:etah1i}).
The complexity of $\kappa$ for $\beta\not=0$ implies that one of
the two frequencies has a positive imaginary part which, in turn,
means that the respective mode is unstable at however small $\beta$.
For $\beta/k^{2}=\varepsilon\ll1$, which is close to the instability
threshold and opposite to the limit of $\beta\gg1$ producing the
so-called edge waves \citep{Morris2003}, we have $\kappa\approx-ik(\varepsilon+i\varepsilon^{2})$
and $\omega\approx\pm k(1+i\varepsilon^{3})$. These relations describe
a slightly oblique wave with the transverse wavenumber $\beta/k\ll k$
which decays over the distance $\sim k^{3}/\beta^{2}$ from the wall
and grows exponentially in time at the rate $\sim\beta^{3}/k^{5}$.
Namely, in this model, the interface can be destabilised by an arbitrary
weak electromagnetic effect which gives rise to a nearly transverse
wave travelling along the wall.

It is instructive to contrast the instability in the semi-infinite
model considered above with that in the slightly more realistic model
of a finite-width channel. This model was first considered by \citet{Davidson1998}
and further studied in the context of hydromagnetic edge waves by
\citet{Morris2003} and \citet{Molokov2011}.

For a channel bounded by two lateral walls at $y=\pm1$, the general
solution of (\ref{eq:etah},\ref{eq:phih}) can be written as 
\begin{eqnarray}
\hat{\eta}(y) & = & \hat{\eta}_{-}\sin(\kappa y)+\hat{\eta}_{+}\cos(\kappa y),\label{gs:etah2}\\
\hat{\phi}(y) & = & \hat{\phi}_{-}\sinh(ky)+\hat{\phi}_{+}\cosh(ky)+\omega^{-2}\hat{\eta}(y),\label{gs:phih2}
\end{eqnarray}
where $\kappa$ is the transverse wavenumber satisfying the dispersion
relation of pure interfacial gravity waves 
\begin{equation}
\omega^{2}=k^{2}+\kappa^{2}.\label{eq:dsp}
\end{equation}

\begin{figure}
\begin{centering}
\includegraphics[width=0.5\columnwidth]{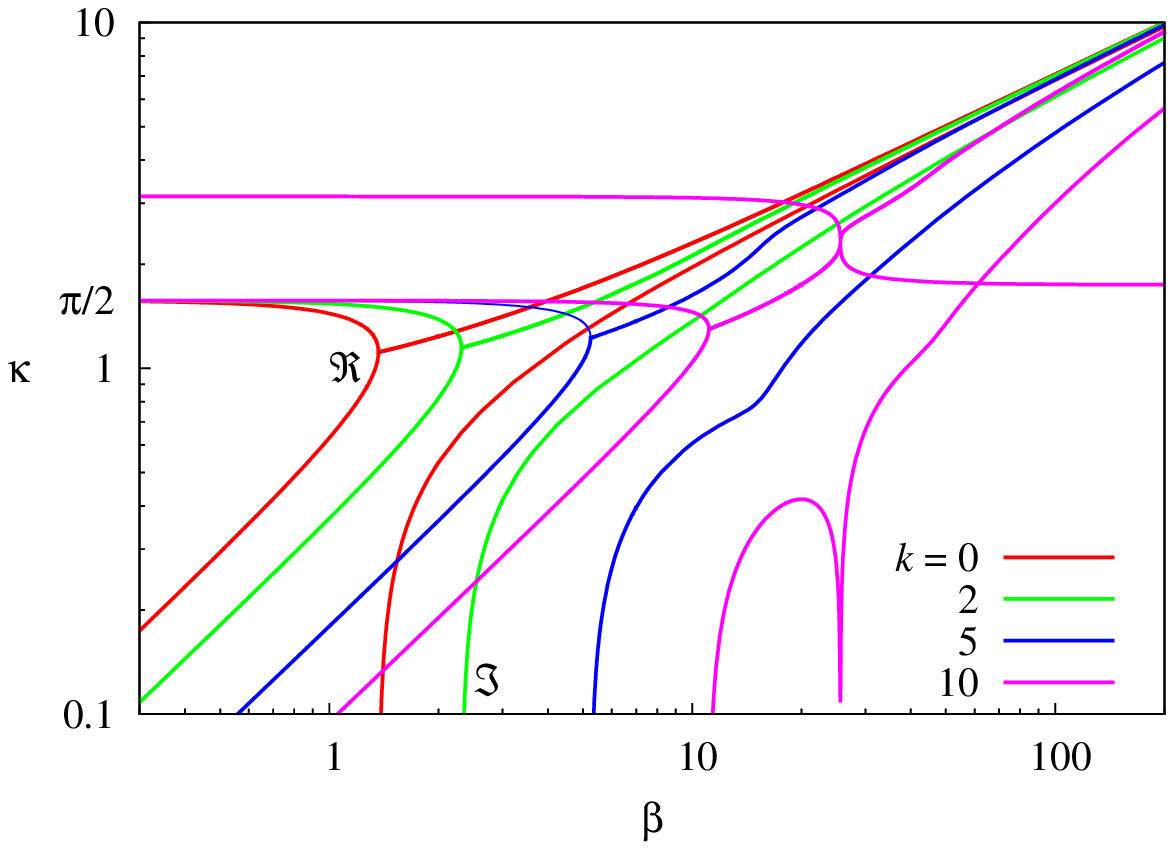}\put(-10,25){(a)}\includegraphics[width=0.5\columnwidth]{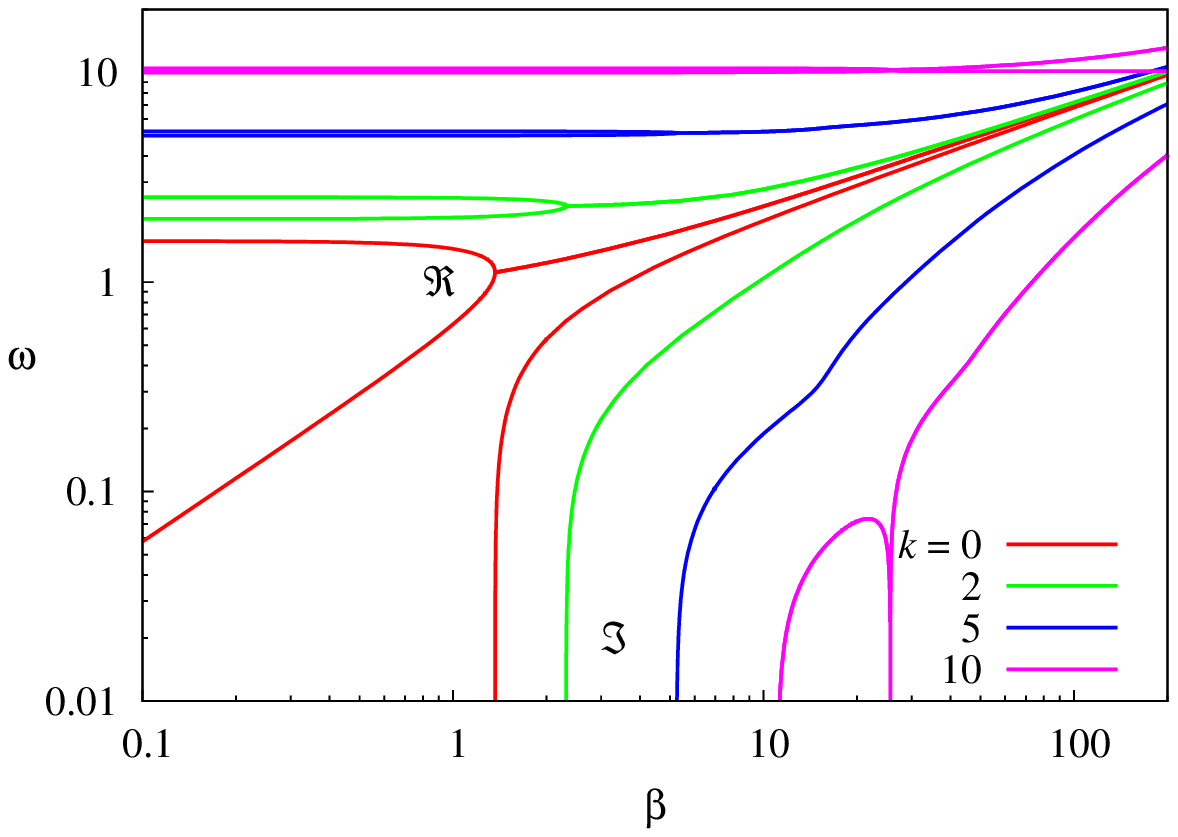}\put(-10,25){(b)}\\
 \includegraphics[width=0.5\columnwidth]{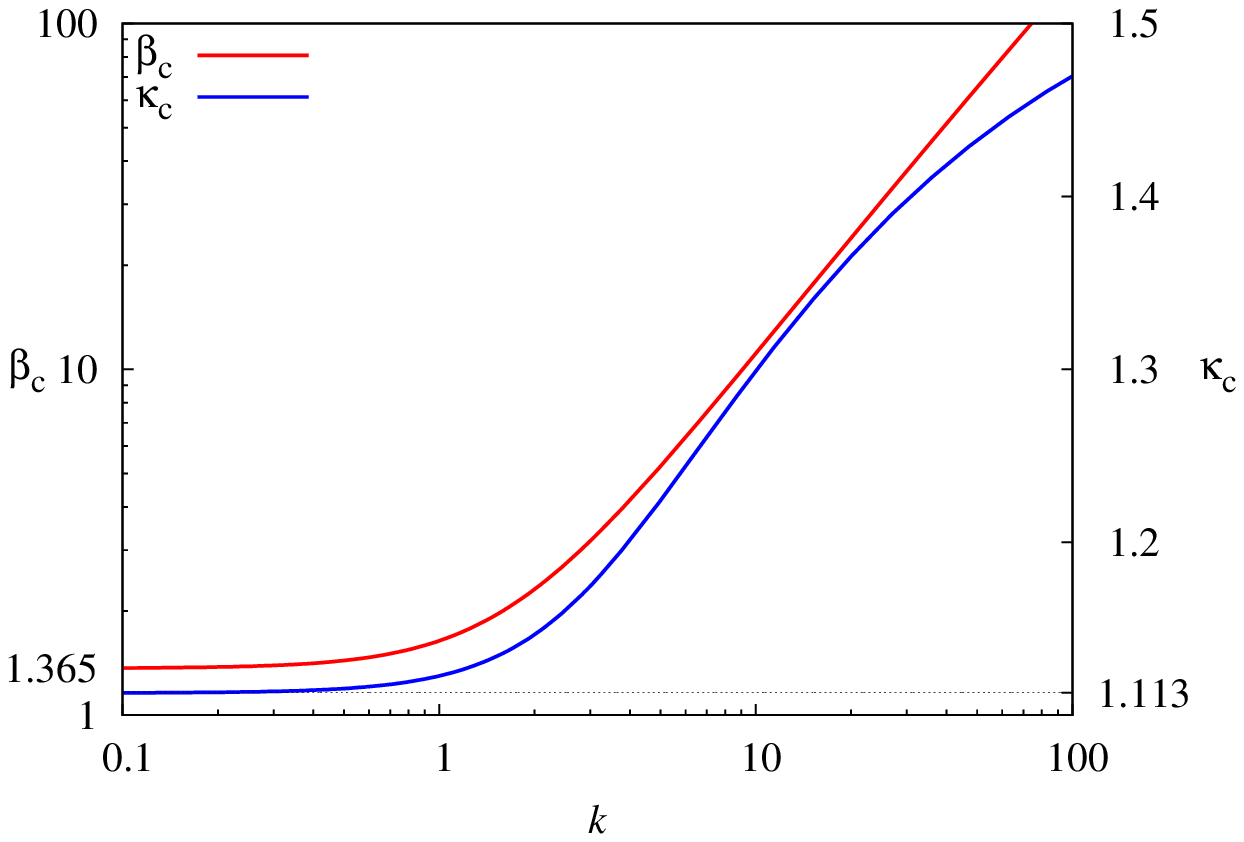}\put(-10,25){(c)}
\par\end{centering}
\caption{\label{fig:kapomg}(a) Real $(\Re)$ and imaginary $(\Im)$ parts
of the first three transverse wavenumbers $\kappa$, (b) the respective
frequencies $\omega$ versus the electromagnetic interaction parameter
$\beta$ for various longitudinal wavenumbers $k$ and (c) the critical
$\beta$ and transverse wavenumbers $\kappa$ as functions of the
longitudinal wavenumbers $k.$}
\end{figure}

Applying boundary conditions (\ref{bc:etah}) at $y=\pm1$, after
a few rearrangements, we obtain 
\begin{equation}
\beta(k,\kappa)=\pm\left(k^{2}+\kappa^{2}\right)\left(\frac{k}{\kappa}\left(\frac{k}{\kappa}+\frac{\tan\kappa}{\tanh k}-\frac{\tanh k}{\tan\kappa}\right)-1\right)^{-1/2},\label{eq:beta}
\end{equation}
which explicitly defines $\beta$ depending on the longitudinal and
transverse wavenumbers $k$ and $\kappa.$ It is important to note
that the respective eigenmode is a pure gravity wave as long as $\kappa$
is real. On the other hand, (\ref{eq:beta}) implicitly defines the
spectrum of admitted $\kappa$ values for given $k$ and $\beta$.
For $\beta=0$, this equation reduces to $\sin2\kappa=0$, which describes
the standard wavenumber spectrum for a channel of dimensionless width
equal to 2: $\kappa_{n}=n\pi/2,\,n=0,1,2,\ldots$. As seen in figure
\ref{fig:kappa}, where $\beta(k,\kappa)$ is plotted against $\kappa$
for various $k$, the increase of $\beta$ just modifies the spectrum
of the wave modes admitted by the electromagnetic reflection condition
(\ref{bc:eta}). It is important to note that the eigenmodes remain
pure gravity waves satisfying the dispersion relation (\ref{eq:dsp})
with real wavenumbers up to the point where two branches of $\kappa$
merge together. The absence of coupling between the gravity wave modes
below the critical point is due to the effective reduction of the
electromagnetic force to the boundary condition (\ref{bc:eta}) which
affects only the reflection but not the propagation of gravity waves.
At the critical point, a complex conjugate pair of wavenumbers emerges,
which means instability as the frequency becomes complex (see figure
\ref{fig:kapomg}). As found by \citet{Morris2003}, who use the reciprocal
of $\beta$, this happens first at $\beta_{c}\approx1.365$ when two
purely longitudinal $(k=0)$ gravity wave modes with $\kappa_{c}=1.113$
merge.

Note that this critical mode is longitudinal and emerges at finite
$\beta_{c}$, whereas the one predicted by the semi-infinite model
is transverse and emerges at $\beta_{c}=0$. In the channel, the latter
mode is recovered in the short-wave limit $k\gg1$. For such waves,
the marginal interaction parameter and the respective transverse wavenumber
can be seen in figure \ref{fig:kapomg}(c) to scale as $\beta\sim k$
and $\kappa\sim1$. The interaction parameter based on the wave length,
which is the relevant horizontal length scale in this limit, then
scales as $\widetilde{\beta}=\beta/k^{2}\sim k^{-1}\rightarrow0$
so recovering the result of the semi-infinite model.

\section{\label{sec:appB}Chebyshev collocation method}

To solve the eigenvalue problem posed by (\ref{eq:etah2}) and (\ref{eq:phih2})
numerically we use a collocation method with the two-dimensional Chebyshev-Lobatto
grid:

\[
(x_{m},y_{n})=\frac{1}{2}\left(\sqrt{\alpha}\cos\frac{\pi m}{M+1},\frac{1}{\sqrt{\alpha}}\cos\frac{\pi n}{N+1}\right),(0,0)\le(m,n)\le(M+1,N+1),
\]
on which the discretised solution $\{\hat{\eta},\hat{\phi}\}(x_{m},y_{n})=\{\tilde{\eta},\tilde{\phi}\}_{m,n}$
and its derivatives are sought. The latter are expressed in terms
of the former by substituting the first and second derivatives with
the differentiation matrices \citep[pp. 393--394]{Peyret2013}. Requiring
the equations to be satisfied at the inner collocation points $(1,1)\le(m,n)\le(M,N)$,
we obtain a system of discrete equations which can be written as 
\begin{eqnarray}
-\lambda\tilde{\eta}_{0} & = & \underline{A}_{0}\tilde{\eta}_{0}+\underline{A}_{1}\tilde{\eta}_{1},\label{eq:eta0}\\
-\tilde{\eta}_{0} & = & \underline{A}_{0}\tilde{\phi}_{0}+\underline{A}_{1}\tilde{\phi}_{1},\label{eq:phi0}
\end{eqnarray}
where the subscripts $0$ and $1$ denote the parts of the solution
at the inner and boundary collocation points respectively. The matrices
$\underline{A}_{0}$ and $\underline{A}_{1}$ represent parts of the
discretised Laplacian using the respective sets of points. The boundary
conditions are applied the boundary points $m=0,M+1$ and $n=0,N+1$,
which results in 
\begin{eqnarray}
\underline{B}_{0}\tilde{\eta}_{0}+\underline{B}_{1}\tilde{\eta}_{1} & = & \beta\underline{C}\tilde{\phi}_{1},\label{bc:eta0}\\
\underline{B}_{0}\tilde{\phi}_{0}+\underline{B}_{1}\tilde{\phi}_{1} & = & \tilde{0},\label{bc:phi0}
\end{eqnarray}
where $\underline{B}$ and $\underline{C}$ are the discretised normal
and tangential derivative operators at the boundary. The above system
of discrete equations is reduced to a matrix eigenvalue problem as
follows. First, (\ref{bc:eta0}) and (\ref{bc:phi0}) can solved for
$\tilde{\phi}_{1}$ and $\tilde{\eta}_{1}$ in terms of $\tilde{\phi}_{0}$
and $\tilde{\eta}_{0}$ as 
\begin{eqnarray}
\tilde{\phi}_{1} & = & -\underline{B}_{1}^{-1}\underline{B}_{0}\tilde{\phi}_{0},\label{eq:phi1}\\
\tilde{\eta}_{1} & = & -\underline{B}_{1}^{-1}\underline{B}_{0}\tilde{\eta}_{0}+\beta\underline{B}_{1}^{-1}\underline{C}\tilde{\phi}_{1}.\label{eq:eta1}
\end{eqnarray}
Substituting (\ref{eq:phi1}) into (\ref{eq:phi0}), we have $\tilde{\eta}_{0}=-\underline{\widetilde{A}}_{0}\tilde{\phi}_{0}$,
where $\underline{\widetilde{A}}_{0}=\underline{A}_{0}-\underline{A}_{1}\underline{B}_{1}^{-1}\underline{B}_{0}$.
Combining this and (\ref{eq:phi1}) with (\ref{eq:eta1}) and (\ref{eq:eta0}),
after a few rearrangements, we obtain the sought matrix eigenvalue
problem

\[
\lambda\tilde{\phi}_{0}=\left[-\underline{\widetilde{A}}_{0}+\beta\underline{\widetilde{A}}_{0}^{-1}\underline{A}_{1}\underline{B}_{1}^{-1}\underline{C}\underline{B}_{1}^{-1}\underline{B}_{0}\right]\tilde{\phi}_{0},
\]
which is solved using the standard DGEEV routine of LAPACK.

\bibliographystyle{jfm}
\bibliography{cell}

\end{document}